\documentclass[twocolumn,amsmath,floatfix,prb,aps]{revtex4}

\usepackage{color}
\usepackage{amsmath}
\usepackage{pifont}   % Ding symbols
\usepackage{graphicx} % Include figure files
\usepackage{dcolumn}  % Align table columns on decimal point
\usepackage{bm}       % bold math
\usepackage{amsfonts} % Some more fonts
\usepackage{amssymb}  % More symbols
\usepackage{multirow} % Table functions
\usepackage{romannum}

\begin{document}
%\DeclareGraphicsRule{*}{png}{*}{}

%%%% User-defined commands %%%%
\newcommand{\ba}{{\bf a}}
\newcommand{\BB}{{\bf b}}
\newcommand{\bd}{{\bf d}}
\newcommand{\br}{{\bf r}}
\newcommand{\bp}{{\bf p}}
\newcommand{\bk}{{\bf k}}
\newcommand{\bg}{{\bf g}}
\newcommand{\bj}{{\bf j}}
\newcommand{\bt}{{\bf t}}
\newcommand{\bv}{{\bf v}}
\newcommand{\bu}{{\bf u}}
\newcommand{\bq}{{\bf q}}
\newcommand{\bG}{{\bf G}}
\newcommand{\bP}{{\bf P}}
\newcommand{\bJ}{{\bf J}}
\newcommand{\bK}{{\bf K}}
\newcommand{\bL}{{\bf L}}
\newcommand{\bR}{{\bf R}}
\newcommand{\bS}{{\bf S}}
\newcommand{\bT}{{\bf T}}
\newcommand{\bQ}{{\bf Q}}
\newcommand{\bA}{{\bf A}}
\newcommand{\bH}{{\bf H}}

\newcommand{\bra}[1]{\left\langle #1 \right |}
\newcommand{\ket}[1]{\left| #1 \right\rangle}
\newcommand{\braket}[2]{\left\langle #1 | #2 \right\rangle}
\newcommand{\mel}[3]{\left\langle #1 \left| #2 \right| #3 \right\rangle}

\newcommand{\bdel}{\boldsymbol{\delta}}
\newcommand{\bsig}{\boldsymbol{\sigma}}
\newcommand{\beps}{\boldsymbol{\epsilon}}
\newcommand{\bnu}{\boldsymbol{\nu}}
\newcommand{\bnab}{\boldsymbol{\nabla}}
\newcommand{\bGam}{\boldsymbol{\Gamma}}

\newcommand{\bgt}{\tilde{\bf g}}

\newcommand{\brh}{\hat{\bf r}}
\newcommand{\bph}{\hat{\bf p}}

\author{R. Gupta$^1$}
\author{S. Maisel$^2$}
\author{F. Rost$^1$}
\author{D. Weckbecker$^1$}
\author{M. Fleischmann$^1$}
\author{H. Soni$^3$}
\author{S. Sharma$^4$}
\author{A. G\"orling$^2$}
\author{S. Shallcross$^1$}
\email{sam.shallcross@fau.de}
\affiliation{1 Lehrstuhl f\"ur Theoretische Festk\"orperphysik, Staudtstr. 7-B2, 91058 Erlangen, Germany.}
\affiliation{2 Lehrstuhl f\"ur Theoretische Chemie, Egerlandstr. 3, 91058 Erlangen, Germany.}
\affiliation{3 School of Sciences, Indrashil University, India.}
\affiliation{4 Max-Born-Institute for non-linear optics, Max-Born Strasse 2A, 12489 Berlin, Germany.}

\title{Deformation induced pseudo-magnetic fields in complex carbon architectures}
\date{\today}

\begin{abstract}
We show that the physics of deformation in $\alpha$-, $\beta$-, and $6,6,12$-graphyne is, despite their significantly more complex lattice structures, remarkably close to that of graphene, with inhomogeneously strained graphyne described at low energies by an emergent Dirac-Weyl equation augmented by strain induced electric and pseudo-magnetic fields. To show this we develop two continuum theories of deformation in these materials: one that describes the low energy degrees of freedom of the conical intersection, and is spinor valued as in graphene, and one describing the full sub-lattice space. The spinor valued continuum theory agrees very well with the full continuum theory at low energies, showing that the remarkable physics of deformation in graphene generalizes to these more complex carbon architectures. In particular, we find that deformation induced pseudospin polarization and valley current loops, key phenomena in the deformation physics of graphene, both have their counterpart in these more complex carbon materials.
\end{abstract}

\maketitle

\section{Introduction}

One of the most remarkable features of graphene is the robustness of its ultra-relativistic low energy physics. The Dirac-Weyl Hamiltonian that describes the quasi-particles of pristine graphene remains a valid description even under substantial deformation, simply being augmented by effective pseudo-magnetic and electric fields that encode the deformation in a low energy description. These fields, except for the requirement that the pseudo-magnetic field change sign at conjugate valleys, behave exactly as physical electric and magnetic fields\cite{gui10,Electronic}, resulting in a rich phenomenology of deformation induced physics in single layer graphene. For realistic strain the induced magnetic field can reach  hundreds of Tesla, a remarkable effect observable as deformation induced Landau levels in graphene\cite{Science-NB}. This deep connection between structural deformation and an induced electromagnetic field promises a control over electronic properties unrivaled in any three dimensional material, and generates novel physical effects such as deformation induced valley filters\cite{Valley-filter, Valley-filter2, Valley-filter6, Valley-filter7, Valley-filter8}, and psuedospin polarization\cite{pseudo1,pseudo2}.

Following the experimental realization of graphene low energy conical intersections have been predicted for several all-carbon materials, each with a substantially more complex lattice structure than that of graphene\cite{6, 5,7,8,13,s1,s2,s5,s7,s8,s9,s11,s12,s13,s14,16,17,23,24,25,26,27,42,46,53,54,61, 43,r1,r2,1}. For example, $6,6,12$-graphyne and $\beta$-graphyne both possess 18 carbon atoms in their unit cell, as opposed to the 2 atom unit cell honeycomb lattice of graphene. This entails a much more difficult chemistry of their fabrication\cite{5,9,44}, but also a fundamental difference in the low energy physics. In graphene sub-lattice space is isomorphic to $SU(2)$ pseudo-spin space, and it is this that underpins the connection between deformation and effective electric and magnetic fields. However, this simple relation between the pseudospin and sub-lattice degrees of freedom is lost in these more complex materials. While the structural physics of non-uniform deformation in the graphynes has been quite intensively investigated\cite{s8,s9,s11,s13}, the corresponding attention has not been devoted to the electronic theory of general non-uniform deformations, with most electronic investigations focusing either on uniform uniaxial and biaxial strains\cite{s1,s2,s5,s12,s14,49,50}, ``rotating'' strain\cite{s7}, or phononic excitations\cite{s4}. A natural question is therefore: how much of the rich electronic physics of non-uniform deformation in graphene finds a counterpart in these more complex carbon architectures?

The purpose of the present paper is to answer this question. To that end we generalize the continuum theory of 
deformation in graphene to materials with arbitrary numbers of atoms in the unit cell. As a minimal description of the electronic structure entails one $\pi$-orbital per basis atom, this theory now necessarily takes two forms: one involving all sub-lattice degrees of freedom, and a down-folded theory describing only the spinor degree of freedom of the low energy Dirac cone.
For all three graphynes we find that this latter description is, at low energies, in very close agreement with the full continuum theory. Thus the intimate connection between structure and effective electromagnetic field is preserved in these more complex architectures, and the rich physics of deformation in graphene generalizes to the semi-metallic graphynes.

\section{Continuum theory of complex carbon materials}

\begin{figure}
  \centering
  \includegraphics[width=0.98\linewidth]{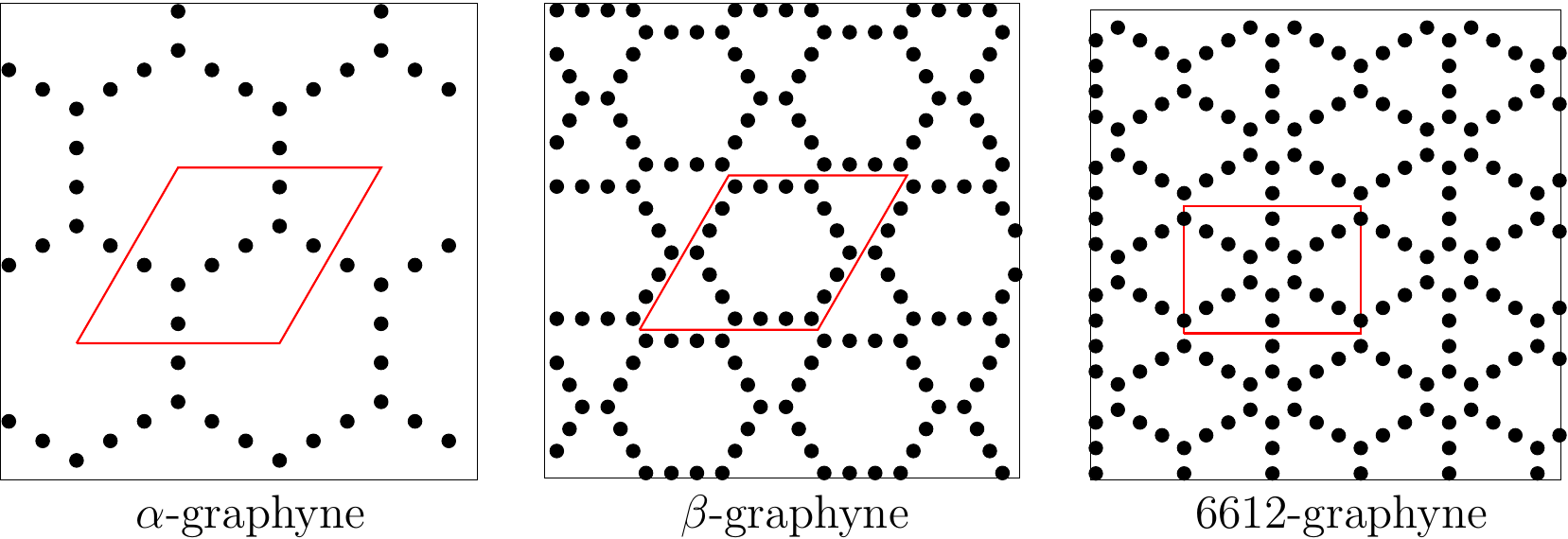}
  \caption{Lattice structures of $\alpha$-graphyne, $\beta$-graphyne, and $\gamma$-graphyne, with the unit cell indicated by the red (light shaded) box.
  }
  \label{F1}
\end{figure}

While the continuum theory of deformation in graphene is very well developed, the same attention has not been paid to the graphynes. The complex lattice structures of these materials (see Fig.~\ref{F1}) render either cumbersome (the expansion of a tight-binding Hamiltonian) or inapplicable (transport of the Dirac-Weyl equation to non-Minkowski metric) the methods used in deriving the continuum theory of deformation for graphene. To avoid this we employ a methodology\cite{kiss15,shall2016,shall17,Gupta2018,Rost2019} based on an exact map of the tight-binding Hamiltonian to a general continuum operator, recently used to treat stacking deformations in bilayer structure\cite{kiss15,shall17}, and acoustic and optical deformations in single layer graphene\cite{Gupta2018}. We first briefly review this theory, before extending it to include a treatment of currents, and subsequently describing both the complete and down-folded versions required to investigate deformation in graphyne.

\subsection{An exact map of the tight-binding Hamiltonian}

In this approach the ``input'' consists simply of a hopping function envelope $t_{\alpha\beta}(\br,\bdel)$ describing the tight-binding hopping matrix amplitude from a position $\br$ on sub-lattice $\alpha$ to $\br+\bdel$ on sub-lattice $\beta$ in the material, with the change in electron hopping due to deformation encoded in the $\br$-dependence of this function. The tight-binding Hamiltonian is therefore

\begin{eqnarray}
\label{tb0}
H_{TB} & = & \sum_{\alpha\bR_i\beta\bR_j} t_{\alpha\beta}(\bR_i+\bnu_\alpha,\bR_j+\bnu_\beta-\bR_i-\bnu_\alpha) \nonumber \\
&& \times c_{\bR_j+\bnu_\beta \beta}^\dagger c_{\bR_i+\bnu_\alpha \alpha}.
\end{eqnarray}
where $\bR_i$ and $\bnu_\alpha$ are the lattice and basis vectors of the underlying high symmetry lattice.

The corresponding continuum Hamiltonian $H(\br,\bp)$ that is exactly equivalent to this tight-binding Hamiltonian is given by\cite{Rost2019}

\begin{equation}
 \left[H(\br,\bp)\right]_{\alpha\beta} = \frac{1}{V_{UC}}\sum_{j} M_{j\alpha\beta} 
 \eta_{\alpha\beta}(\br,\bK_j+\bp/\hbar),
 \label{HM}
\end{equation}
where $V_{UC}$ is the unit cell volume, the sum is over all $\bG_j$ vectors of the reciprocal lattice, $\bK_j = \bK_0 + \bG_j$ with $\bK_0$ a reference momentum in the Brillouin zone (for the graphynes we study this will be the momentum at which the conical intersection occurs), 

\begin{equation}
\eta_{\alpha\beta}(\br,\bq) = \int d\bdel\,e^{-i\bq.\bdel} t_{\alpha\beta}(\br,\bdel)
\label{MSH}
\end{equation}
the so-called mixed space hopping function, and the $M_j$ matrices are given by

\begin{equation}
 M_{j\alpha\beta} = e^{i\bG_j.(\bnu_\alpha-\bnu_\beta)}
\end{equation}
As shown in Ref.~\onlinecite{Gupta2018} expansion of the $\br$-dependence of $\eta_{\alpha\beta}(\br,\bq)$ for slowly varying deformations and expansion of the $\bp$-dependence for momenta near the conical intersection leads to a systematic treatment of deformation within the continuum picture.

\subsection{An exact map of the current operator}

In this work a focus will be on the valley current densities induced by deformation, and thus we require a current operator corresponding to the Hamiltonian Eq.~\ref{HM}. The current operator in tight-binding theory, the so-called bond current operator, is given by\cite{tod02,boy10}

\begin{equation}
 \bj(\bR_i) = \frac{1}{V_{UC}} \sum_{\alpha}\left[ n(\bR_i + \bnu_{\alpha}) \bv + \bv n(\bR_i + \bnu_{\alpha}) \right]
 \label{jTB}
\end{equation}
where $\hat{n}(\bR_i + \bnu_{\alpha}) = \ket{\bR_i+\bnu_{\alpha}}\bra{\bR_i+\bnu_{\alpha}}$ is the density operator with the velocity operator

\begin{equation}
 {\bf v} = \frac{1}{i\hbar}\left[\br,H\right],
\end{equation}
and where we have included normalization by the unit cell volume $1/V_{UC}$.

One would expect that the method used to derive Eq.~\eqref{HM} from Eq.~\eqref{tb0} should, if applied to Eq.~\eqref{jTB}, yield a current density given by

\begin{equation}
 \bj(\br) = \Im\Psi(\br)^\dagger\left[\bnab_\bp H(\br,\bp) \Psi(\br)\right]
 \label{j}
\end{equation}
with $H(\br,\bp)$ given by Eq.~\eqref{HM}, i.e. in the continuum limit 
the relation between Hamiltonian and current operator should follow from Hamilton's equations. However, given the very different forms of the tight-binding Hamiltonian Eq.~\eqref{tb0} and the bond-current operator Eq.~\eqref{jTB} in tight-binding theory it is not obvious that this is the case.

Indeed, the precise link between the tight-binding and continuum limits of the current operator has been the subject of recent discussion in the context of recovering the Sch\"odinger current operator from the bond current formulae\cite{boy10} for a quadratic band model. As noted in Ref.~\onlinecite{boy10} the fundamental difficultly involves taking the continuum limit of a discrete grid and the associated ambiguity in the definition of differential operators. Employing the same methodology involved in the derivation of Eq.~\ref{HM}, however, we find a general form for the continuum current operator precisely equivalent to the general tight-binding bond-current formula while avoiding all use of grid limits. This turns out to be just the intuitive result given by Eq.~\eqref{j} from which we can then recover, as specific cases, both the Schr\"odinger and Dirac-Weyl current operators.

To derive this result we first consider the matrix element of Eq.~\ref{jTB} with a general state of the system

\begin{equation}
 \ket{\Psi} = \sum_{\bk_1\alpha} c_{\bk_1\alpha} \ket{\Phi_{\bk_1\alpha}}
\end{equation}
where $\ket{\Phi_{\bk_1\alpha}}$ denotes a Bloch function of sublattice $\alpha$ and crystal momentum $\bk_1$:

\begin{equation}
 \ket{\Phi_{\bk_1\alpha}} = \frac{1}{\sqrt{N}} \sum_{\bR_i} e^{i \bk_1.(\bR_i+\bnu_\alpha)}\ket{\bR_i+\bnu_\alpha}
 \label{Bloch}
\end{equation}
One finds for the matrix element $\mel{\Psi}{\bj(\bR_i)}{\Psi}$ the result

\begin{eqnarray}
&& \frac{1}{2 V_{UC}} \sum_{\bk_1,\bk_2,\alpha',\beta} \Big[c_{\bk_1\alpha'}^\ast c_{\bk_2\beta} \mel{\Phi_{\bk_1\alpha'}}{\hat{n}(\bR_i + \nu_{\alpha}) \hat{{\bf v}}}{\Phi_{\bk_2\beta}} \nonumber \\
&& + \text{h.c.} \Big] 
\label{eqstart}
\end{eqnarray}
We now work out the matrix element $\mel{\Phi_{\bk_1\alpha'}}{\hat{n}(\bR_i + \nu_{\alpha}) \hat{{\bf v}}}{\Phi_{\bk_2\beta}}$ in detail. Insertion of the Bloch functions, Eq.~\ref{Bloch}, yields

\begin{eqnarray}
&&\frac{-i}{N\hbar} \sum_{\bR_j} e^{-i\bk_1.(\bR_i + \bnu_\alpha)}  e^{i\bk_2.(\bR_j+\bnu_\beta)} \left[\bR_i+\bnu_\alpha-\bR_j-\bnu_\beta\right] \nonumber \\
&&\times t_{\alpha\beta}(\bR_i+\bnu_\alpha,\bR_j+\bnu_\beta)
\label{xyz}
\end{eqnarray}
where $t_{\alpha\beta}(\bR_i+\bnu_\alpha,\bR_j+\bnu_\beta) = \mel{\bR_i+\bnu_\alpha}{H}{\bR_j+\bnu_\beta}$ is the usual tight-binding hopping matrix element. To derive a continuum limit we now employ the Poisson sum formula in the form

\begin{equation}
 \sum_{\bR_j} f(\bR_j+\bnu_\beta) = \frac{1}{V_{UC}}\sum_{\bG_j} \hat{f}(\bG_j) e^{i\bG_j.\bnu_\beta}
 \label{PS}
\end{equation}
for which the appropriate function $f(\br)$ is

\begin{equation}
 f(\br) = e^{i\bk_2.\br}\left(\bR_i+\bnu_\alpha-\br\right) t_{\alpha\beta}(\bR_i+\bnu_\alpha,\br)
\end{equation}
with the Fourier transform

\begin{equation}
 \hat{f}(\bq) = \int d\br\, e^{-i(\bq-\bk_2).\br}\left(\bR_i+\bnu_\alpha-\br\right) t_{\alpha\beta}(\bR_i+\bnu_\alpha,\br)
\end{equation}
Evaluation of the integral and employing the Poisson sum in Eq.~\eqref{xyz} then yields

\begin{equation}
\frac{e^{i(\bk_2-\bk_1).(\bR_i+\bnu_\alpha)}}
 {V\hbar} \sum_{\bG_j} e^{i\bG_j.(\bnu_\alpha-\bnu_\beta)}\bnab_\bq t_{\alpha\beta}(\bR_i + \bnu_\alpha,\bk_2+\bG_j)
\end{equation}
which, by defining a reference momenta $\bK_0$ through $\bk_2+\bG_j = \bG_j + \bK_0 + \bp_2$ and promotion of $\bp_2$ to an operator, can straightforwardly be recast into a form involving the continuum Hamiltonian $H(\br,\bp)$ Eq.~\eqref{HM}:

\begin{equation}
 \frac{1}{\sqrt{V}} e^{-i\bp_1.\br} 
 \left\{ \bnab_\bp H_{\alpha\beta}(\br,\bp) \right\}
 \frac{1}{\sqrt{V}} e^{i\bp_2.\br}.
 \label{mel-compact}
\end{equation}
To make the connection with a continuum Hamiltonian we introduce the vector plane waves

\begin{eqnarray}
\phi_{\bp_1\alpha}(\br) & = & \frac{1}{\sqrt{V}} e^{i\bp_1.\br} \ket{\alpha} \\ 
\phi_{\bp_2\beta}(\br) & = & \frac{1}{\sqrt{V}} e^{i\bp_2.\br}\ket{\beta},
\end{eqnarray}
where $\ket{\alpha}$ and $\ket{\beta}$ represent unit vectors in a space of dimension equal to the number of atomic degrees of freedom (for graphene these would just be pseudospin up and pseudospin down states). Employing these functions we then arrive at the desired operator equivalence:

\begin{equation}
  \mel{\Phi_{\bk_I\alpha}}{\hat{n}(\br) \hat{{\bf v}}}{\Phi_{\bk_J\beta}} = \phi_{\bp_1\alpha}^\dagger(\br) \left[\bnab_\bp H(\br,\bp) \phi_{\bp_2\beta}(\br)\right]
\end{equation}
Insertion of this result back into Eq.~\ref{eqstart} and the obvious definition for the system wavefunction in the continuum representation as

\begin{equation}
 \Psi(\br) = \sum_{\bp_1} c_{\bp_1\alpha} \phi_{\bp_1\alpha}(\br)
\end{equation}
then leads to our final result

\begin{equation}
\Psi(\br)^\dagger \bj \Psi(\br) = \frac{1}{2}\left\{\Psi(\br)^\dagger\left[\bnab_\bp H \Psi(\br)\right]
+ \left[\bnab_\bp H \Psi(\br)\right]^\dagger \Psi(\br)\right\}
\label{jm}
\end{equation}
which is evidently the sought for intuitive form given by Eq.~\eqref{j}.

This expression trivially reproduces both the well known current operators for the Dirac-Weyl and Schr\"odinger equation. For a Schr\"odinger form $H = \frac{1}{2m} p^2$ we find

\begin{equation}
 \bj(\br) = \frac{1}{2m}\left[\Psi(\br)^\ast \bp \Psi(\br) - \Psi(\br) \bp \Psi(\br)^\ast\right]
 \label{Sj}
\end{equation}
whereas for the Dirac-Weyl Hamiltonian $H = v_F \bsig.\bp$ we have

\begin{equation}
 \bj(\br) = \Psi(\br)^\dagger v_F \bsig \Psi(\br)
 \label{DWj}
\end{equation}
There is, however, an important caveat to Eq.~\eqref{jm}. While the bond current operator always satisfies a discrete form of the continuity equation appropriate for the tight-binding Hamiltonian\cite{tod02} the continuum version of the bond current, 
Eq.~\eqref{jm}, is not guaranteed to satisfy the continuity equation. This follows as it is simply the expectation value of the velocity operator. Curiously, it turns out that Eq.~\eqref{jm} indeed violates the continuity equation, but only for Hamiltonians containing a higher than second power in momentum\cite{CE}, a common occurrence in the effective Hamiltonians of condensed matter. Other less intuitive definitions have been provided\cite{Correct-current-operator1,Correct-current-operator2,definition} and these can encode non-classical current contributions. However the breakdown of a classical relationship between velocity and current should probably be viewed as a failure of effective Hamiltonian theory.

\subsection{The continuity equation for deformation in Dirac-Weyl materials}

As our focus here will be on deformation in materials with low energy conical intersections, and the valley currents that deformation induces, it is useful to demonstrate that the current operator defined in the previous section indeed satisfies the continuity equation for such systems. To this end we consider the most general form of the Dirac-Weyl equation augmented by deformation induced fields which, up to second order in momentum, is given by

\begin{eqnarray}
 H &=& A_i(\br) \sigma_i + i \Gamma_i(\br) \sigma_i + v_F^{ij}(\br) \sigma_i p_j  + i w_F^{ij}(\br) \sigma_i p_j \nonumber \\
 & + & \frac{1}{2!} M^{ijk}(\br) \sigma_i p_j p_k
 \label{defH}
\end{eqnarray}
In this expression there are two effective gauge field terms: $\bA$, a real valued gauge\cite{2002-paper,ref13,PhysRevLett.108.227205,non-uniform-strain, AMORIM20161,Peeters-revisited} that transforms in a complex way under spatial rotations, and $\bGam$, an imaginary gauge\cite{PhysRevLett.108.227205} that transforms as a field term in the Dirac-Weyl operator.
$v_F^{ij}$ and $w_F^{ij}$ are the real \cite{PhysRevLett.108.227205,non-uniform-strain,AMORIM20161,Peeters-revisited} and imaginary velocity tensors, in which the $i$ index runs over the three Pauli matrices $\sigma_0$, $\sigma_1$, and $\sigma_2$ and the $j$ index over the two degrees of freedom of space. Finally there is also a trigonal warping term $M^{ijk}$ in which again $i$ runs over the Pauli matrices and the indices $j,k$ over the two degrees of spatial freedom. Higher orders of momenta can be included in Eq.~\eqref{defH} but, if Hermiticity is to be preserved, only by restricting the spatial fields to be slowly varying\cite{Gupta2018}.

It is not immediately evident that Eq.~\eqref{defH} satisfies the continuity equation $\bnab.\bj(\br) + \partial_t n(\br) = 0$ (as both the Hamiltonians $v_F \bsig.\bp$ and $v_F \bsig.\bp + \sigma_i A_i$ obviously do) due both to the presence of the coordinate dependent velocity and mass tensors, as well as the fact that both gauge and velocity terms have both real and imaginary parts. However, as we now show using conditions that guarantee hermiticity of the deformation Hamiltonian, the continuity equation is indeed satisfied.

The $j$'th component of the corresponding current operator is given by

\begin{equation}
j_j = V_F^{ij}(\br)\sigma_i + i W_F^{ij}(\br) \sigma_i + M^{ijk}(\br)\sigma_i p_j.
\end{equation}
with the divergence of the current density then given by

\begin{eqnarray}
\bnab .\bj(\br)&=&\frac{1}{i\hbar}\Im\Bigl[(p_j\Psi)^{\dagger} M^{ijk}\sigma_i p_j \Psi - \psi^{\dagger} M^{ijk}\sigma_i p_j p_k \Psi\nonumber\\
&+&
 i(p_j\Psi)^{\dagger} W^{ij}\sigma_i\Psi- i\psi^{\dagger} W^{ij}\sigma_i p_j\Psi\nonumber\\
&+&  (p_j\Psi)^{\dagger} V^{ij}\sigma_i\Psi - \Psi^{\dagger} V^{ij}\sigma_i p_j \Psi\nonumber\\
&-&\Psi^{\dagger} p_j M^{ijk}\sigma_i p_j \Psi -\hbar\Psi^{\dagger} \partial_j W^{ij}\sigma_i\Psi\nonumber\\
&-&\Psi^{\dagger} p_j V^{ij}\sigma_i\Psi\Bigr].
\end{eqnarray}
Using the obvious relations
$\Im[(p_j\Psi)^{\dagger} M^{ijk}\sigma_i p_j \Psi]=0$ and 
$(\psi^{\dagger} V^{ij}\sigma_i p_j\Psi)^{\dagger}= (V^{ij}\sigma_i p_j\Psi)^\dagger\Psi$ 
in conjunction with the hermiticity conditions obeyed by the Hamiltonian\cite{PhysRevLett.108.227205,Gupta2018}

\begin{eqnarray}
p_j M^{ijk}\sigma_i p_k - 2 i  W^{ij}\sigma_ip_j& =& 0,\\
p_j V^{ij}\sigma_i - 2i \Gamma_i \sigma_i & =& 0,
\end{eqnarray}
we have

\begin{equation}
\bnab .\bj(\br) = \frac{1}{i\hbar}\Im\Bigl[-2\Psi^{\dagger}H\Psi \Bigr]
\end{equation}
and using

\begin{equation}
 \partial_t n(\br) = \frac{1}{i\hbar}\left[\Psi^\dagger\left(H\Psi\right) - \left(H\Psi\right)^\dagger\Psi\right]
\end{equation}
we then find

\begin{equation}
\partial_t n(\br) + \bnab.\bj(\br)=0,
\end{equation}
and so the the continuity equation is satisfied.

\begin{figure*}[tbp]
  \centering
  \includegraphics[width=0.98\linewidth]{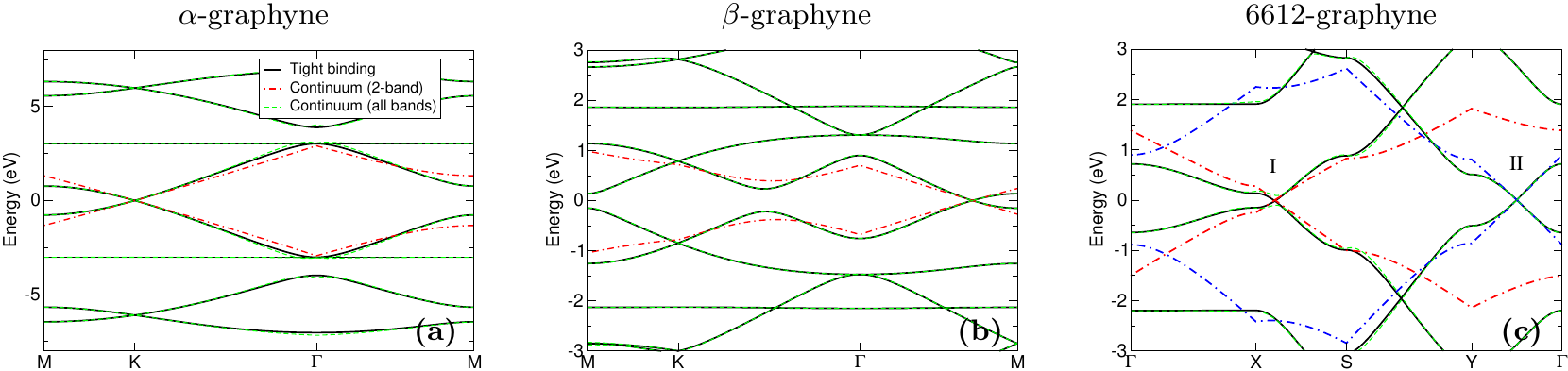}
  \caption{Tight-binding band structures (full line) of (a) $\alpha$-graphyne (b) $\beta$-graphyne and (c) $\gamma$-graphyne. Shown also are the band structure generated by two continuum approximations; one designed to reproduce the low energy Dirac-Weyl conical intersection (dot-dashed lines) and one designed to reproduce all bands (dashed lines). Both these continuum theories include only up to second order in momentum.  In panel (c) the expansion point for the all-band continuum description is the Dirac point of cone II although, as can be observed, this generates a reasonable agreement even close to cone I.
  }
  \label{F2}
\end{figure*}

\section{Continuum theory of $\alpha$-,$\beta$-, and $6,6,12$-graphyne}

\subsection{Tight-binding parameterization}
\label{TBparam}

A continuum theory of the pristine lattices, i.e. without deformation, is easily obtained from Eq.~\eqref{HM}. In this case the mixed space hopping function, Eq.~\eqref{MSH}, loses its $\br$-dependence becoming simply the Fourier transform of the electron hopping function defined between each sub-lattice: $\hat{t}_{\alpha\beta}(\bq) = \int d\bdel\, e^{-i\bq.\bdel} t_{\alpha\beta}(\bdel)$. We therefore require a functional form of $t_{\alpha\beta}(\bdel)$ describing electron hopping for the three graphynes we consider: $\alpha$-, $\beta$-, and $6,6,12$-graphyne. To obtain such a form we take the tight-binding parameters of Ref.~\onlinecite{liu12}, which are defined only for the nearest neighbour hopping, and fit to a Gaussian functional form

\begin{equation}
 t(\bdel^2) = A \exp (-B \bdel^2)
 \label{Thop}
\end{equation}
such that the nearest neighbour hopping is reproduced, with negligible next nearest (and further) hopping. %For $6,6,12$-graphyne, however, we find adjustment of the parameterization is necessary to reproduce the behaviour of the Dirac cone under deformation.

This Gaussian form is useful as (i) it renders the Fourier transform straightforward and (ii) deformation, which modifies the hopping matrix element through changes of the hopping vector, is easily included. The tight-binding band structures using this parameterization are shown as the full lines in Fig.~\ref{F2}. The expected low energy conical intersections are found: (i) at the K-point for $\alpha$-graphyne, (ii) on the line connecting the $\Gamma$ and M-points for $\beta$-graphyne, and (iii) two low energy cones one at the X-point (we denote this cone I) and a second on the line connecting $\Gamma$- and Y-points (denoted cone II) for $6,6,12$-graphyne. In fact, cone I is shifted somewhat further off the X-point as compared to \emph{ab-initio} calculations (a line fraction of 0.2 as opposed to 0.08), and cone II somewhat closer to the $\Gamma$ point (a line fraction of 0.5 as opposed to 0.62 in \emph{ab-initio}). Furthermore, with only nearest neighbour hopping the cone tilting found in \emph{ab-initio} for $\beta$-graphyne and cone II in $6,6,12$-graphyne is not reproduced, lowering the value of $B$ somewhat in Eq.~\eqref{Thop} we find restores the cone tilting. We have checked that sensible variation of the tight-binding parameters does not significantly change the results we present in subsequent sections for deformation in these materials.

\subsection{Continuum theory for pristine lattices}

To extract a tractable continuum description from Eq.~\eqref{HM} requires a Taylor expansion in momentum about the Dirac point. Expanding the hopping function in Eq.~\eqref{HM} to first order

\begin{equation}
 \hat{t}_{\alpha\beta}(\bK_j+\bp) \approx \hat{t}_{\alpha\beta}(\bK_j^2) + \bp.\bnab_\bq \left.\hat{t}_{\alpha\beta}(\bq^2)\right|_{\bq=\bK_j}
\end{equation}
generates an expression of the form

\begin{equation}
 H_0^{full}(\bp) = H^{(0)} + H_x^{(1)} p_x + H_y^{(1)} p_y
 \label{HF}
\end{equation}
where the Hamiltonian at the Dirac momenta is

\begin{equation}
H^{(0)}_{\alpha\beta} = \frac{1}{V_{UC}} \sum_j M_{\alpha\beta j} \hat{t}_{\alpha\beta}^{(0)}(\bK_j^2)
\label{1}
\end{equation}
and the matrices $H_i^{(1)}$ given by

\begin{equation}
 \left[H^{(1)}_i\right]_{\alpha\beta} = \frac{2}{V_{UC}} \sum_j M_{\alpha\beta j} \hat{t}_{\alpha\beta}^{(1)}(\bK_j^2) K_{ji}
 \label{2}
\end{equation}
where $K_{ji}$ is the $i$'th component of the vector $\bK_j$.
The matrices Eqs.~\eqref{1}-\eqref{2} are labelled by sub-lattice indices, i.e. represent $8\times8$ matrices for $\alpha$-graphyne and $18\times18$ matrices for $\beta$-graphyne and $6,6,12$-graphyne. These therefore describe not only the low energy conical intersection (described generically by a spinor degree of freedom) but also all other  band manifolds. In Fig.~\ref{F2} is shown the band structure obtained from Eq.~\eqref{HF}, with additionally second order in momentum terms included, revealing an excellent agreement with the underlying tight-binding method.

To obtain a description of the low energy conical intersection we must down-fold the full continuum theory. To that end we diagonalize the Hamiltonian at the Dirac point $H^{(0)}$, Eq.~\eqref{1}, and apply the resulting unitary transformation $U$ to the full Hamiltonian $H^{full}_0(\bp)$. This yields

\begin{equation}
 H = \beps + U H^{(1)}_x U^\dagger p_x + U H^{(1)}_y U^\dagger p_y
\end{equation}
where $\beps$ is a diagonal matrix whose entries are the eigenvalues at the Dirac point momenta, with the matrices $H^{(1)}$ now encoding hybridization of these bands at a finite momentum away from the Dirac point. This now allows us to identify the sub-space corresponding to degenerate eigenvalues at the Dirac point, and by retaining only this sub-space one arrives at a spinor valued low energy Hamiltonian. This procedure yields a Dirac-Weyl form, but with an SU(2) rotation. Undoing this with a further unitary transform $U_s$ we then find a generic final form for all three materials\cite{16,46}

\begin{equation}
H_0(\bp) = v_x \sigma_x p_x  + v_y \sigma_y p_y + (t_x p_x + t_y p_y)\sigma_0
\label{HC}
\end{equation}
The band structure calculated using Eq.~\eqref{HC} is shown as the dot-dashed lines in Fig.~\ref{F2}, showing good agreement with the full tight-binding calculation for the low energy conical intersection. Further improvement requires additional bands to be included in the continuum theory, not higher orders of momenta.

While Eq.~\eqref{HC} provides a good description of the low energy manifold it does not directly provide the physical wavefunction; for this a back transformation is required. Given a spinor eigenvector $\phi$ of Eq.~\eqref{HC} we must firstly transform back to the global SU(2) frame

\begin{equation}
 c = U_s \phi
 \label{t1}
\end{equation}
which then provides the coefficients for constructing the physical wavefunction from the Dirac point wave-functions:

\begin{equation}
 \Psi_i = \sum_{j\in low} c_j U_{ij}
 \label{t2}
\end{equation}
where $\Psi_i$ is the $i$'th component of the physical wavefunction (i.e., in sub-lattice space) while the sum $j$ is over the band indices of the low energy manifold (recall that $U$ is the unitary transform that diagonalizes $H_0$ the Hamiltonian at the Dirac point momenta). In graphene such a back transformation is, of course, unnecessary; a low energy expansion directly yields a Dirac-Weyl equation. For any more complex material, however, the coefficients of the Dirac-Weyl spinor wavefunction merely parameterize the low energy conical intersection in terms of the Dirac point wavefunctions.

\subsection{Continuum theory of deformation}

Inclusion of deformation into the low energy scheme requires describing how electron hopping changes throughout the material, i.e. the full function $t_{\alpha\beta}(\br,\bdel)$. This can be obtained simply from the geometric information of how the deformation changes the hopping $\bdel$ at point $\br$ in the material, via substitution of $\bdel$ in the hopping function of the high symmetry material $t_{\alpha\beta}(\bdel)$ by $t_{\alpha\beta}(\bdel(\br))$. The change in the square of the hopping vector due to the applied deformation field $\bu(\br)$ is just

\begin{equation}
 \bdel^2 \to \left(\bdel + \bu(\br+\bdel) - \bu(\bdel)\right)^2
\end{equation}
which, upon substitution into a hopping function $t_{\alpha\beta}(\bdel^2)$ yields,  via a Taylor expansion for slowly varying fields, the mixed space hopping function

\begin{eqnarray}
  \eta_{\alpha\beta}(\br,\bq)
 & = & \hat{t}_{\alpha\beta}^{(0)}(\bq^2) \\
 &+& t_{\alpha\beta}^{(1)}(\bq^2)\left(\epsilon_{xx}(\br)+\frac{1}{2}(\partial_x \bu(\br))^2\right) q_x^2 \nonumber\\ 
 &+& \hat{t}_{\alpha\beta}^{(1)}(\bq^2)\left(\epsilon_{yy}(\br)+\frac{1}{2}(\partial_y \bu(\br))^2\right) 
 q_y^2 \nonumber \\
 &+& \hat{t}_{\alpha\beta}^{(1)}(\bq^2) \left(2\epsilon_{xy}(\br)+\partial_x \bu(\br).\partial_y \bu(\br)\right)q_x q_y \nonumber
\end{eqnarray}
In this expression $t_{\alpha\beta}^{(n)}(\bq^2)$ are the the Fourier transforms of derivatives of the high symmetry hopping function:

\begin{equation}
 t_{\alpha\beta}^{(n)}(\bq^2) = \int \!\!d\bdel e^{i\bdel.\bq} \frac{\partial^n t(\bdel^2)}{\partial{(\bdel^2)}^n}
\end{equation}

Insertion of this result directly into Eq.~\eqref{HM} leads to a Hamiltonian of the form

\begin{eqnarray}
 H_{def}^{full} & = & H_{xx} \left(\varepsilon_{xx} + \frac{1}{2} (\partial_x\bu)^2\right)
 + H_{yy} \left(\varepsilon_{yy} + \frac{1}{2} (\partial_y\bu)^2\right) \nonumber \\
 &+& H_{xy} \left(2\epsilon_{xy}(\br)+\partial_x \bu(\br).\partial_y \bu(\br)\right) 
 \label{HDF}
\end{eqnarray}
with the matrices $H_{nm}$ given by

\begin{figure}[tbp]
  \centering
  \includegraphics[width=0.6\linewidth]{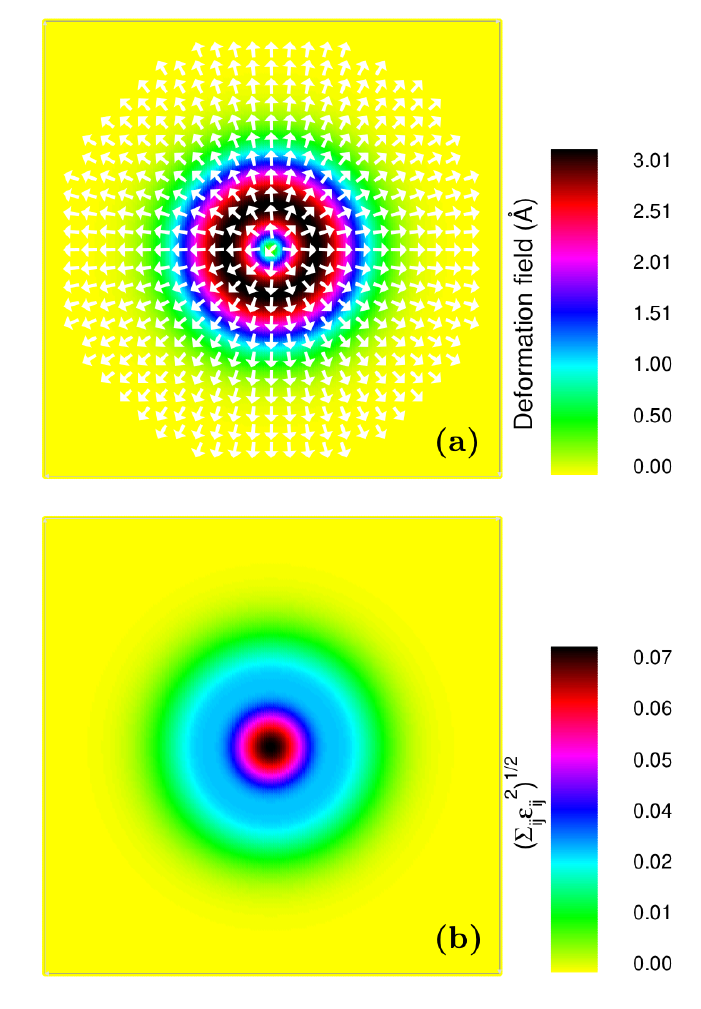}
  \caption{The in-plane deformation applied to all materials; the length scale of the unit cell is 100a in all cases. In panel (a) is shown the deformation field with the colour bar the magnitude and the arrow the direction, while in panel (b) is shown the magnitude of the deformation tensor $\epsilon$.
  }
  \label{F3}
\end{figure}

\begin{equation}
 \left[H_{nm}\right]_{\alpha\beta} = \frac{1}{V_{UC}} \sum_j M_{\alpha\beta j} \hat{t}_{\alpha\beta}^{(0)}(\bK_j^2) K_{jn}K_{jm}
\end{equation}
where again $K_{jn}$ is the $n$'th component of the vector $\bK_j = \bK_0 + \bG_j$. This along with Eq.~\ref{HF} provides a continuum description in which all bands are included, i.e., in terms of all sub-lattice degrees of freedom.

To identify the low energy sector of the Hamiltonian with deformation, we employ the same transformations $U$ and $U_s$ that yielded the low energy sector for the pristine lattice. This, for all three materials, then results in the following generic form for the low energy deformation Hamiltonian

\begin{eqnarray}
H_{def}(\br,\bp) &=& [\varepsilon_{xx} + \frac{1}{2} (\partial_x\bu)^2] (\sigma_0 f_{xx} + \sigma_x g_{xx})\nonumber \\
&+& [\varepsilon_{yy} + \frac{1}{2} (\partial_y\bu)^2] (\sigma_0 f_{yy} + \sigma_x g_{yy})\nonumber\\
&+& [2\varepsilon_{xy} + \partial_x \bu . \partial_y \bu)](\sigma_y g_{xy})
\label{HD}
\end{eqnarray}
where we have suppressed the $\br$-dependence is the deformation tensor and fields, and where the $f_{ij}$ and $g_{ij}$ are constants.
This is very close to the form of deformation in graphene, which can obtained by substituting $f_{xx} = f_{yy} = \alpha$ and $g_{xx} = - g_{yy} = - g_{xy} = \beta$ into this expression. This latter condition,

\begin{equation}
g_{xx} = -g_{yy} = -g_{xy}
\label{cc}
\end{equation}
will hold for all materials in which the Dirac cone is protected by symmetry, as this ensures that biaxial strain cannot displace the Dirac cone of the high symmetry point. This is true for $\alpha$-graphyne, and we find Eq.~\ref{cc} to be exactly satisfied by our deformation expansion, but not for $\beta$- or $6,6,12$-graphyne. For these materials biaxial deformation shifts the Dirac cone in momentum space; for $\beta$-graphyne towards the X point while for $6,6,12$-graphyne cone I moves towards the $X$ point and cone II towards the $Y$ point\cite{50}.
%This magnitude of this change is given by $g_{xx}+g_{yy}$ and we find this to be somewhat smaller in our effective Hamiltonian scheme than that found in our ab-initio calculations. However, if the atomic positions are frozen, the shift of the cone under biaxial strain is much closer to the tight-binding results. The indicates that local relaxation can be important in these materials, an effect that also occurs in graphene\cite{non-uniform,non-uniform2,Gupta2018}. While this can be included in the formalism we present here, for the case of non-uniform strain that will be our primary interest the cancellation in $g_{xx}+g_{yy}$ of two quantities of approximately equal and opposite magnitude, essential for the cone shift under biaxial strain, is less important.

\begin{figure}[tbp]
  \centering
  \includegraphics[width=0.78\linewidth]{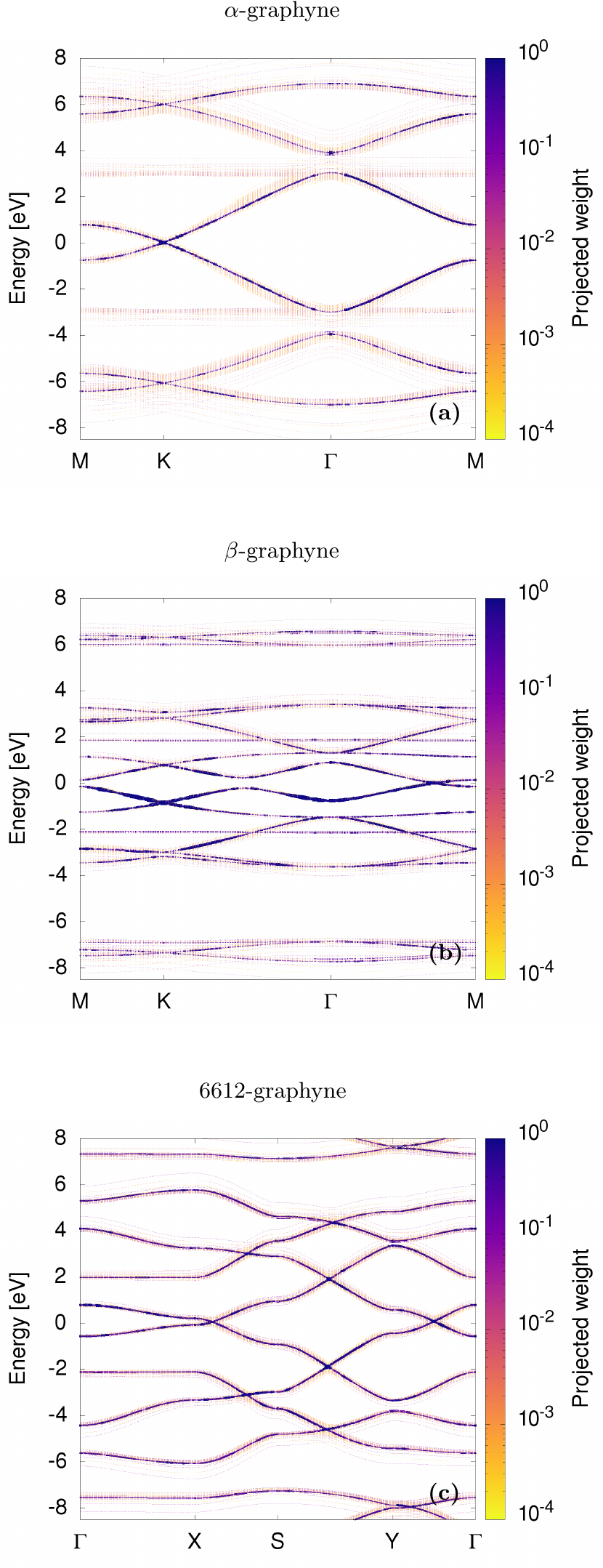}
  \caption{Band manifolds in the extended
zone scheme for $\alpha$-graphyne, $\beta$-graphyne, and $6,6,12$-graphyne. Shown is the projection of the wavefunction of the system with deformation onto the set of wavefunctions of the ideal system at $\bk$, see Eq.~\ref{spec}, with the amplitude of the projection indicated by the colour. These plots therefore represent the broadening of the eigenstates of the pristine material due to scattering induced by the deformation. By comparison with Fig.~\ref{F2}, which displays the band structures of the corresponding systems without deformation, we see that while the low energy conical intersections are broadened by deformation, they are not significantly disrupted.}
  \label{F4}
\end{figure}

\section{Electronic structure of non-uniform deformation in graphyne}

Having established the basic theory we now examine the electronic consequences of non-uniform deformation. After a brief description of the numerical methodology, we first examine the robustness of the low energy conical manifolds under deformation, before considering the changes in electron densities and current densities induced by deformation.

\begin{figure*}[tbp]
  \centering
  \includegraphics[width=0.98\linewidth]{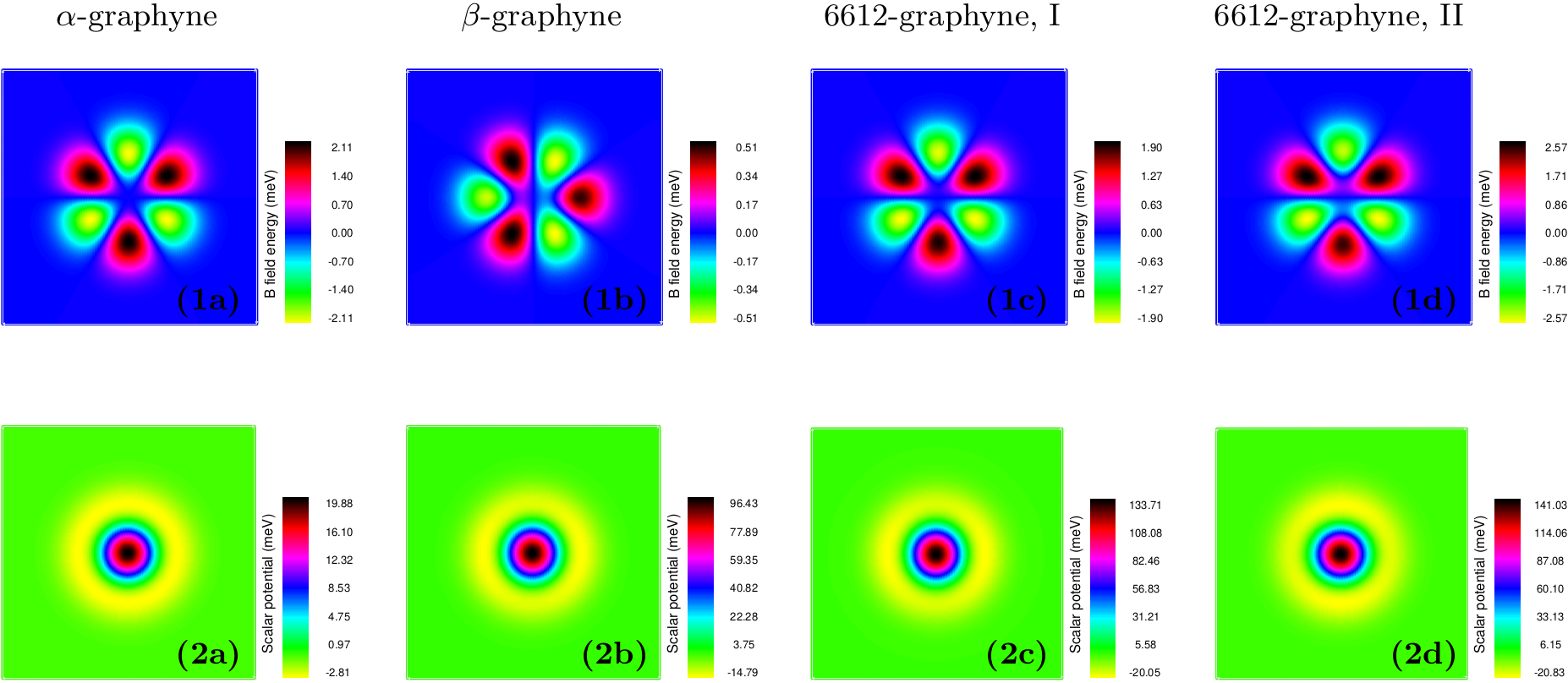}
  \caption{\emph{Deformation induced magnetic fields in the graphynes}: Shown are effective pseudo-magnetic field and scalar potential generated by the deformation field displayed in Fig.~\ref{F3}, for $\alpha$- and $\beta$-graphyne, and the two conical intersections of $6,6,12$-graphyne. For ease of comparison we present the magnetic field as an energy via multipliction of the Fermi velocity of each material. As can be seen, despite the very different lattice structures the effective fields induced by deformation are remarkably similar in form (the 90$^\circ$ rotation in $\beta$-graphyne is due to the choice of Dirac cone).
  }
  \label{F5}
\end{figure*}

\subsection{Numerical details}

Solution of either the 2-band or full band Hamiltonian with deformation, Eqs.~\ref{HDF} and \ref{HD}, is performed using a basis of eigenstates from the pristine system. For a momentum $\bk_0$ the basis set is formed from all eigenstates of the pristine system with momenta $|\bk-\bk_0| < k_{cut}$ and energy $|\epsilon^{(0)}_{\bk i}| < e_{cut}$. The advantage of this basis lies in the efficiency of convergence: we find that typically to converge the electronic structure in an energy window $E$ requires $e_{cut} \approx 1.5 E$. The basis size is determined by $e_{cut}$ with $k_{cut}$ chosen so as the restrict the calculation to a single valley. We find that $e_{cut}$ chosen so that the basis size is of between 1000 and 2000 states is usually sufficient for convergence. For further numerical details we refer the reader to Ref.~\onlinecite{Gupta2018}.

\begin{figure*}[tbp]
  \centering
  \includegraphics[width=0.98\linewidth]{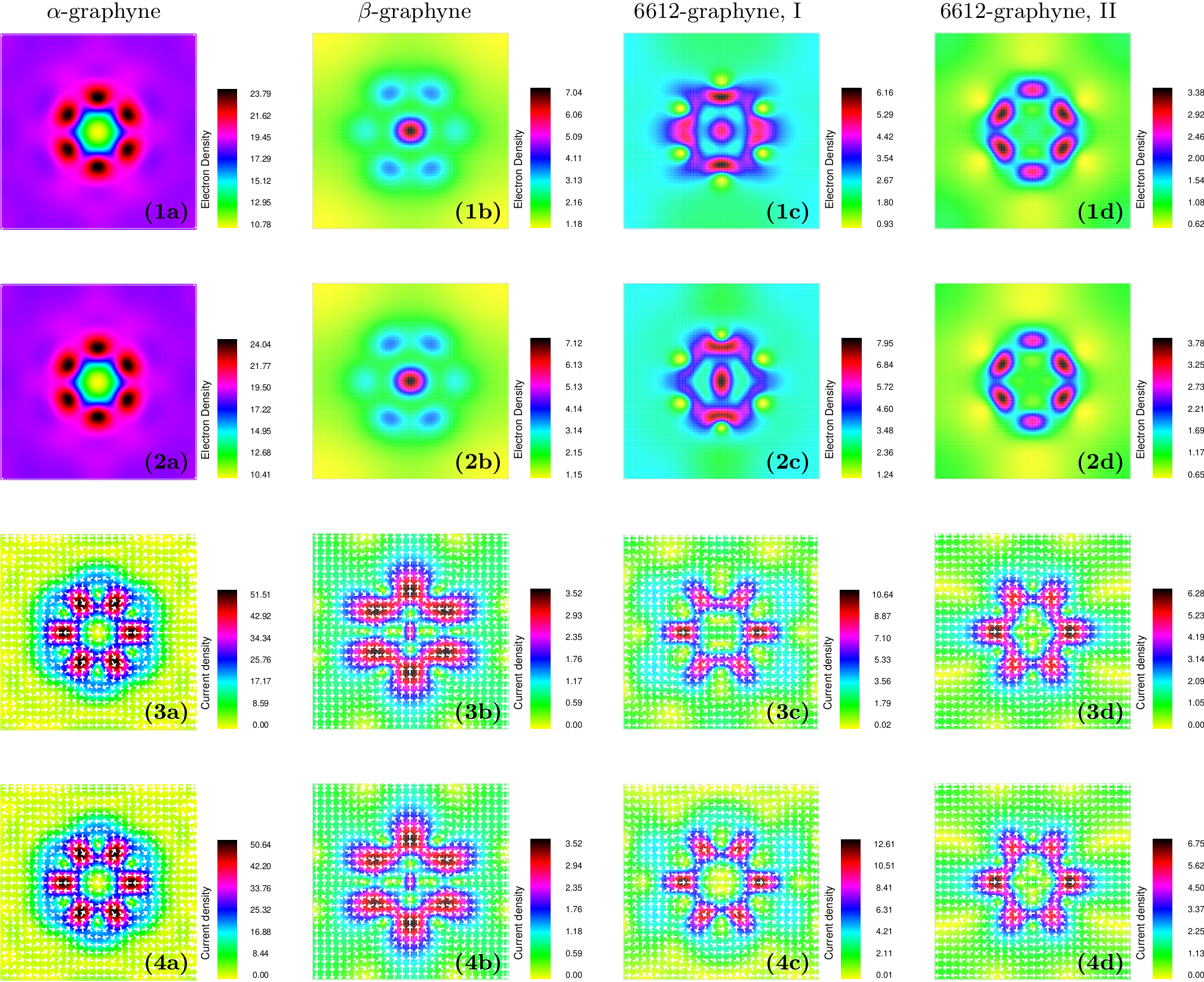}
  \caption{\emph{Density and current density induced by the deformation in the graphynes}. Shown in columns 1 to 4 are results for, respectively,  $\alpha$-graphyne, $\beta$-graphyne, the Dirac cone on the X-S symmetry line of $6,6,12$-graphyne, and the Dirac cone that lies on the $\Gamma$-Y symmetry line of $6,6,12$-graphyne. In each case the deformation field that field in Fig.~\ref{F3}. The energy window within which the density and current density are integrated is, $0$-$100$~meV for $\alpha$-graphyne, $0$-$10$~meV for $\beta$-graphyne, and $20$-$30$~meV at both conical intersections in $6,6,12$-graphyne; similar results are however seen for any low energy window. In the first and second rows are shown the electron density generated by the 2-band and full-band continuum pictures respectively, while in rows 3 and 4 are shown the current density produced by the 2- and full-band continuum theories respectively. In all cases it can be seen that the 2- and full band continuum theories are in excellent agreement, and thus the Dirac-Weyl description of deformation remains valid for these materials, despite their complex lattice structures. While $\alpha$-graphyne exhibits exactly the same ``charge flowers'' found in graphene, see panels (1a) and (2a) for the other graphynes this is not the case with, in particular, cone I of $6,6,12$-graphyne, panels (1c) and (2c), clearly exhibiting the $C_2$ symmetry of the underlying lattice.
  }
  \label{F6}
\end{figure*}

\subsection{Spectral weight changes due to deformation}

How are the band manifolds of the pristine lattices modified by deformation? As large scale periodic deformation results in a Brillouin zone reduced by a factor of $10^2$-$10^3$ as compared to the pristine lattice, the multiply back-folded and hybridized  band structure becomes very hard to interpret. A much more useful quantity is what could be called a ``poor man's spectral function'':

\begin{equation}
 \omega(\bk,\epsilon) = \sum_j \rho_{\bk j} \delta(\epsilon-\epsilon_{\bk j})
 \label{spec}
\end{equation}
where

\begin{equation}
 \rho_{\bk j} = \sum_i \braket{\phi_{\bk i}}{\Psi_{\bk j}}
\end{equation}
In this expression $\ket{\Psi_{\bk j}}$ is the $j$'th eigenstate at crystal momentum $\bk$ of the system with deformation and $\ket{\phi_{\bk i}}$ an eigenstate, also at $\bk$, for the system without deformation. In the absence of deformation $\rho_{\bk j} = 1$ and Eq.~\ref{spec} is simply the band structure of the pristine material. However in the presence of deformation $\ket{\Psi_{\bk j}}$ will involve the coupling together of many eigenstates of the pristine system and $\rho_{\bk j} < 1$. Thus Eq.~\ref{spec} represents how the bands of the high symmetry system are broadened through scattering induced by the deformation. To explore this we consider the deformation field shown in Fig.~\ref{F3}, similar to those typically employed in the discussion of non-uniform deformation in graphene. In panel (a) of this figure is shown the deformation field, while in panel (b) is shown the magnitude of the deformation tensor. The maximum value of the strain tensor is $\sim 7$\%; our ab-initio calculations indicate that strains of $< 7$\% are within the elastic regime of these materials.
As can be seen in Fig.~\ref{F4} for each of the three graphynes we consider the ``spectral function'' follows closely the band structures of the pristine systems (see Fig.~\ref{F2}) but with the expected deformation induced broadening. The non-uniform ``speckled'' nature of the spectral intensity along the band lines can be understood as arising from the complex multiple intersections and subsequent hybridization that occurs when the bands are folded back to the Brillouin zone of the deformed system;  in the extended zone scheme this will be manifest as a non-uniform spectral weight along the band lines.

\subsection{Charge inhomogenity and current flow}

Having established the robustness of the low energy manifold to deformation, we now consider a description of deformation within a continuum theory of the low energy conical intersection. In this case the physics is encoded in the effective electric and pseudo-magnetic fields that augment the Dirac-Weyl equation, and these fields are shown in Fig.~\ref{F5} for $\alpha$-, $\beta$- and $6,6,12$-graphyne, each with the same circularly symmetric deformation field shown in Fig.~\ref{F3}. Strikingly, for all three materials the form of the effective pseudo-magnetic and scalar fields is very similar. This is remarkable when one considers the very different lattice structures of these three systems, with $\alpha$- and $\beta$-graphyne possessing hexagonal lattices  and 6,6,12-graphyne a rectangular lattice. While the pseudo-magnetic fields are comparable in magnitude, the scalar field is almost an order of magnitude greater for cone II of 6,6,12-graphyne than for $\alpha$-graphyne suggesting that the interplay of gauge and scalar fields, known to be significant for describing nanobubbles in graphene, would be especially important in this material\cite{Science-NB,PhysRevB.96.241405}.

\begin{figure}[tbp]
  \centering
  \includegraphics[width=0.98\linewidth]{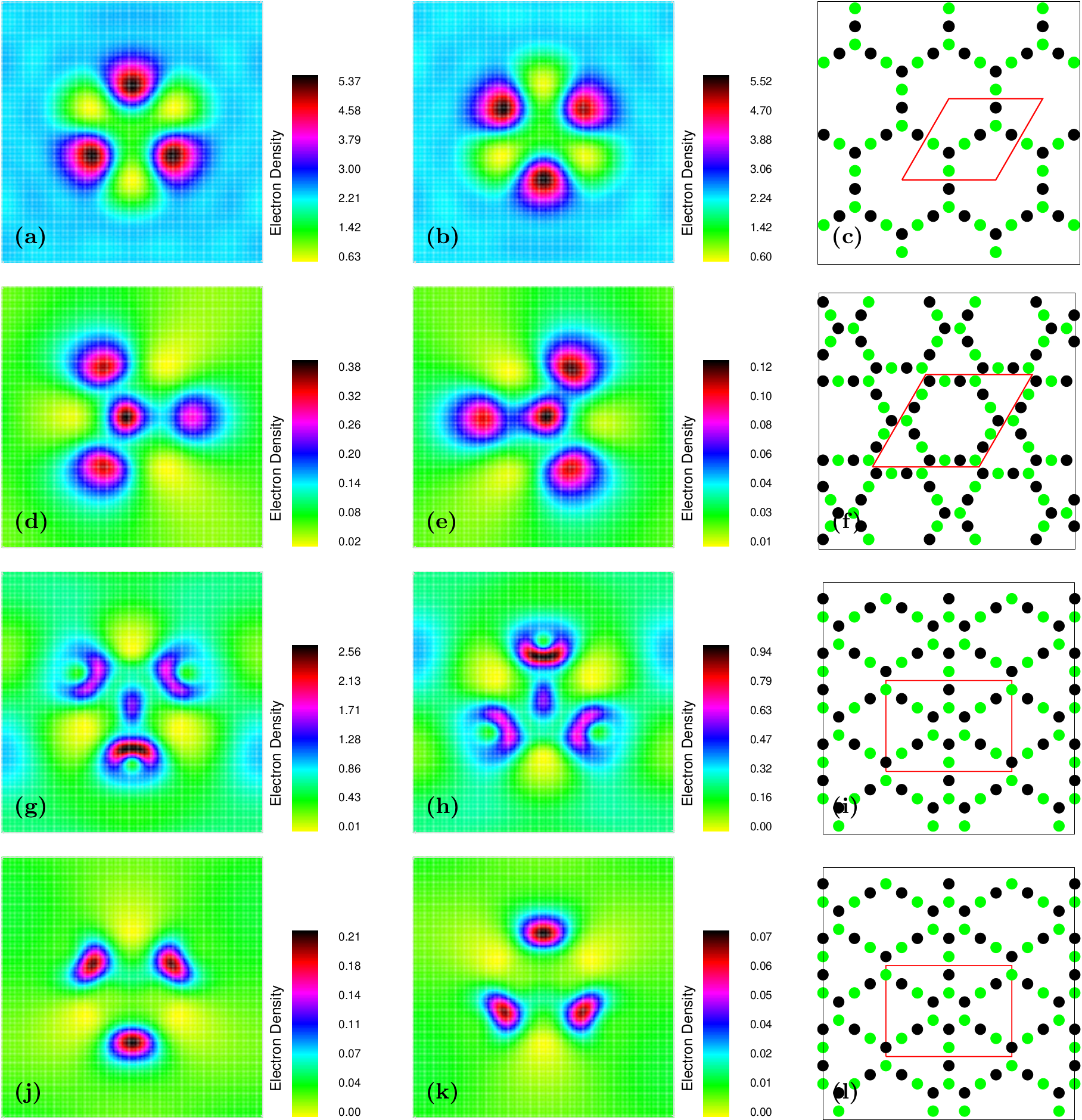}
  \caption{\emph{Pseudospin polarization near the Dirac point}. Shown in each row is the atom projected density for $\alpha$-graphyne, $\beta$-graphyne, and cones I and II of $6,6,12$-graphyne respectively for the same deformation and integration window employed in Fig.~\ref{F6}. For all atoms in the unit cell of these materials the atomic projected electron density takes on one of two forms exhibited in each of the first two columns. The assignment of each atom in the unit cell to each of the two projection types is shown in the third column with the A and B type atoms, corresponding to columns one and two respectively, shown by dark and light (green) shading respectively. Pseudospin polarization due to deformation, well known in graphene, thus generalizes to the graphynes.
  }
  \label{F7}
\end{figure}

\begin{figure}[tbp]
  \centering
  \includegraphics[width=0.98\linewidth]{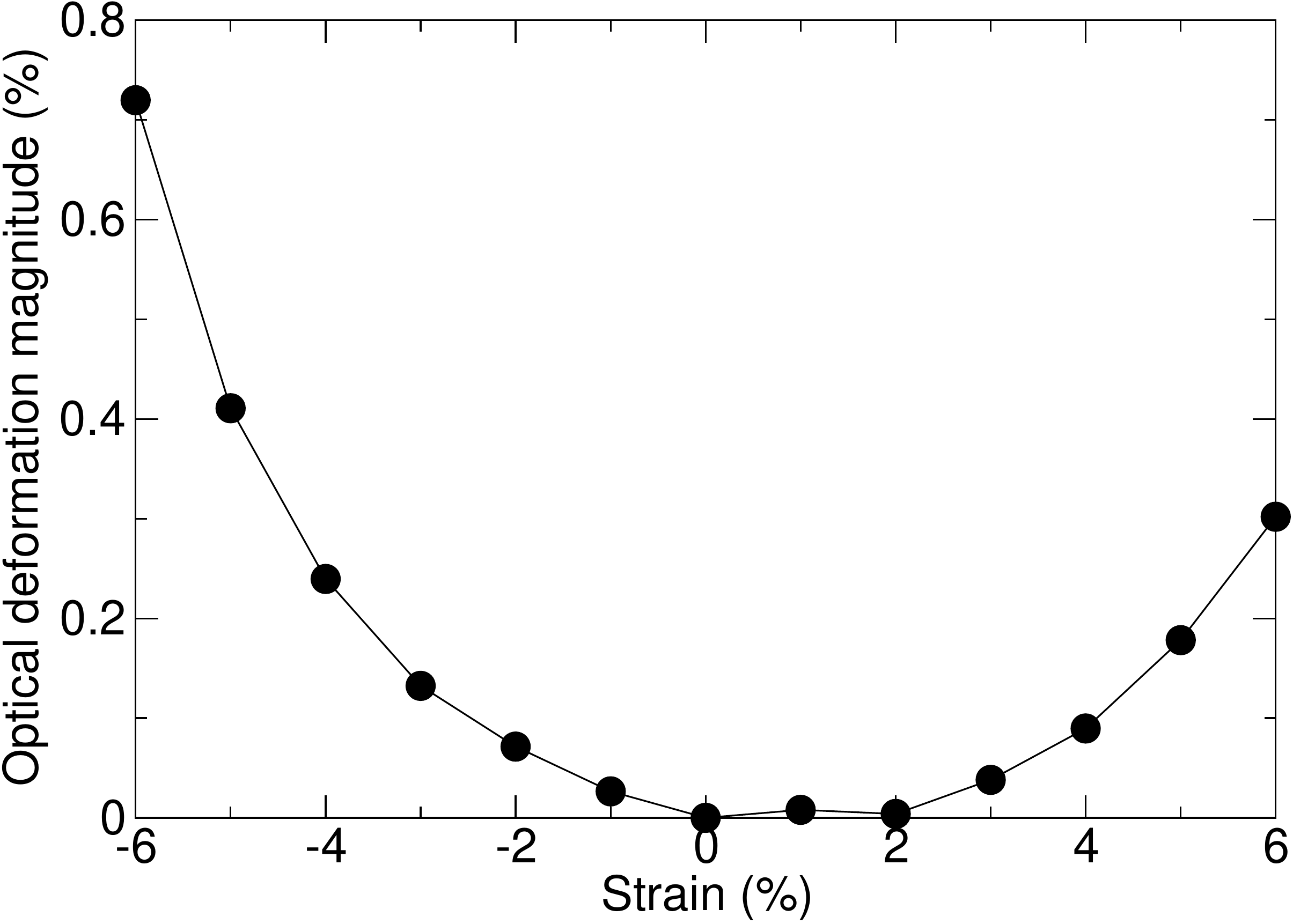}
  \caption{\emph{Optical deformation due to local atomic relaxation induced by biaxial strain in 6,6,12-graphyne}. On each of the groups of atoms on which pseudospin polarization occurs (see Fig.~\ref{F7}) we define the average displacement of the atoms from their ideal positions under strain. From these two average displacements may then be calculated the acoustic and optical components of the deformation, with the magnitude of the latter plotted as a function of applied strain. Note that the acoustic component, as expected, is zero.
  }
  \label{opt}
\end{figure}

Is the 2-band Dirac-Weyl theory of deformation valid in these more complex carbon architectures? To probe this question we now examine the deformation induced changes in electron density and current density calculated using the 2-band and full-band continuum theories. If the Dirac-Weyl continuum theory provides a valid description, then results from these two distinct continuum theories should be in close agreement. In rows 1 and 2 of Fig.~\ref{F6} are shown electron density using the 2-band and full-band continuum theory respectively, with in rows 3 and 4 displayed the current density, again calculated from the 2- and full-band continuum theory respectively. For both density and current density it can be seen that the results of the two continuum theories are in very good agreement: the Dirac-Weyl description of deformation thus remains valid for these much more complex lattice structures.

We now consider the structure of the deformation induced changes to electron density and current density. For $\alpha$-graphyne these are very similar to those reported in previous studies of graphene for circularly symmetric deformation fields, with ``charge flowers''\cite{Gaussian} of $C_6$ symmetry and an associated $C_3$ symmetry current density pattern\cite{eddy_currents,currents2,currents3}, see panels 1a-4a of Fig.~\ref{F6}. For $\beta$- and $6,6,12$-graphyne, however, the density exhibits a much lower symmetry, especially striking for cone I of 6,6,12-graphyne. This reflects the $C_2$ lattice symmetry for 6,6,12-graphyne. The corresponding current densities are, however, much closer in form to those found in $\alpha$-graphyne and graphene.

As the effective fields induced by deformation showed no great distinction of form the differences in electron density indicates the importance of the transformation from the pseudo-spinor of the Dirac-Weyl equation back to the physical wavefunction discussed in Sec. II. However, the deformation induced current densities do correspond to the pseudo-magnetic fields shown in Fig.~\ref{F5}, with regions of strong current flow occurring at the nodal lines of the pseudo-magnetic field. The current density induced by deformation in these graphynes can therefore, just as in graphene, be understood as due to snake states\cite{2000,currents2,currents3} at regions where the pseudo-magnetic field changes sign.

Finally we examine the question of sub-lattice polarization. In graphene the deformation induced charge density is strongly polarized on sub-lattice A or B, which can be viewed in the Dirac-Weyl picture as a local pseudospin polarization. The more complex lattice structures for $\alpha$-, $\beta$-, and $6,6,12$-graphyne, which have 8, 18, and 18 atoms in the unit cell respectively, suggests that this physics will not be transferable to these more complex carbon architectures. Remarkably, as can be seen in Fig.~\ref{F7}, this is not the case. For each of these materials the atom projected density takes on one of only two forms, exhibited in the first two columns, with the correspondence between projection type and atom position in the unit cell shown in column three.
However, while each of the 8 atoms of $\alpha$-graphyne have a density given exactly by one of these two projection types, for $\beta$-graphyne and 6,6,12-graphyne there are slight deviations amongst the 9 atom projections of each type.

\subsection{The role of optical deformation}

Thus far we have not considered the role of local relaxations which will undoubtedly be induced by application of a deformation field. This is known to be more significant in these materials than in graphene (where it can also qualitatively change the physics\cite{midgap}). In the case of graphene the two atom unit cell leads to a natural effective Hamiltonian theory in terms of acoustic and optical deformation fields, as recently discussed by Gupta et al.\cite{Gupta2018}. However, the more complex unit cells of the graphynes imply many more optical modes. To simplify this situation, and by analogy with graphene, we can define optical and acoustic modes in terms of the two groups of atoms on which charge is localized due to pseudospin polarization. In this way we can define the average displacement of each group of atoms off their ideal positions under strain, and so define single optical and acoustic modes.

To investigate this we have performed \emph{ab-initio} calculations for $6,6,12$-graphyne using the VASP software suite, in which we allow the 18 atoms of the unit cell to relax under an applied biaxial strain. In Fig.~\ref{opt} is shown the magnitude of the resulting optical deformation, due to local atomic relaxation, given as a percentage of the nearest neighbour separation for a range of biaxial strains. As can be seen, for biaxial strains of up to 6\%, the optical deformation is of the order of 1\%. Beyond 8\% biaxial strain the lattice substantially reconstructs. Thus optical deformation will likely play some role in the physics of these materials, as they do with graphene, and may in principle be included in an effective Hamiltonian description following the scheme outlined in Ref.~\cite{Gupta2018}.

\section{Discussion}

%We have shown that continuum operators that are the formal operator equivalent to the tight-binding Hamiltonian, $H_{TB}$, and to the bond current expression, $\bj_{TB}(\br)$, are related by Hamilton's equations, as one would intuitively expect. That is to say, if $H_{TB} \leftrightarrow H(\br,\bp)$ and $\bj_{TB}(\br) \leftrightarrow \mel{\Psi}{\bj(\br,\bp)}{\Psi}$, with $\leftrightarrow$ indicating operator equivalence, then $\mel{\Psi}{\bj(\br,\bp)}{\Psi} = \mel{\Psi}{\bnab_\bp H(\br,\bp}{\Psi}$. This generalizes the discussion, for example in Ref.~\onlinecite{boy10} on the relation between the tight-binding and continuum current operators.

The principal question we have addressed is whether the intuitive Dirac-Weyl description of deformation in graphene generalizes to the more complex carbon architectures of the graphynes. To answer this question we have developed two distinct continuum theories: a Dirac-Weyl type theory, formally identical to that of graphene, and an continuum approach describing the full band structure. For $\alpha$-, $\beta$-, and $6,6,12$-graphyne these lead to very similar results for the deformation induced changes to the density and current density close to the Dirac point, showing that the Dirac-Weyl description remains valid for these materials. Deformation in the graphynes thus retains the remarkable connection between structural change and pseudo-magnetic and scalar fields found in graphene, and the rich physics of that material -- valley filters\cite{Valley-filter, Valley-filter2, Valley-filter6, Valley-filter7, Valley-filter8}, pseudospin polarization\cite{pseudo1,pseudo2}, and deformation induced Landau ladders\cite{Science-NB} -- can be expected to be found in the graphynes, if they can be synthesized.

\section*{Acknowledgement}

This work was carried out in the framework of SFB 953 of the Deutsche Forschungsgemeinschaft (DFG).

%\bibliographystyle{unsrt}
%\bibliography{graphyne}

\end{document}